\documentclass[preprint,aps,superscriptaddress]{revtex4}

\usepackage{graphicx}
\usepackage{subfigure}
\usepackage{epsfig}
\usepackage{xcolor}
\usepackage{dcolumn}
\usepackage{bm}
\usepackage{ulem}
\usepackage{color}
\usepackage{epstopdf}

\begin{document}
%\linenumbers

\title{Electronic structure of transferred graphene/h-BN van der Waals heterostructures with nonzero stacking angles by nano-ARPES}

\author{Eryin Wang}
\affiliation{State Key Laboratory of Low Dimensional Quantum Physics and Department of Physics, Tsinghua University, Beijing 100084, China}

\author{Guorui Chen}
\affiliation{State Key Laboratory of Surface Physics and Department of Physics, Fudan University, Shanghai 200433, China}

\author{Guoliang Wan}
\affiliation{State Key Laboratory of Low Dimensional Quantum Physics and Department of Physics, Tsinghua University, Beijing 100084, China}

\author{Xiaobo Lu}
\affiliation{Beijing National Laboratory for Condensed Matter Physics and Institute of Physics, Chinese Academy of Sciences, Beijing 100190, China}

\author{Chaoyu Chen}
\affiliation{Synchrotron SOLEIL, L'Orme des Merisiers, Saint Aubin-BP 48, 91192 Gif sur Yvette Cedex, France}

\author{Jose Avila}
\affiliation{Synchrotron SOLEIL, L'Orme des Merisiers, Saint Aubin-BP 48, 91192 Gif sur Yvette Cedex, France}

\author{Alexei V. Fedorov}
\affiliation{Advanced Light Source, Lawrence Berkeley National Laboratory, Berkeley, CA 94720, USA}

\author{Guangyu Zhang}
\affiliation{Beijing National Laboratory for Condensed Matter Physics and Institute of Physics, Chinese Academy of Sciences, Beijing 100190, China}
\affiliation{Collaborative Innovation Center of Quantum Matter, Beijing, P.R. China}

\author{Maria C.Asensio}
\affiliation{Synchrotron SOLEIL, L'Orme des Merisiers, Saint Aubin-BP 48, 91192 Gif sur Yvette Cedex, France}

\author{Yuanbo Zhang}
\affiliation{State Key Laboratory of Surface Physics and Department of Physics, Fudan University, Shanghai 200433, China}

\author{Shuyun Zhou}
\altaffiliation{Correspondence should be sent to syzhou@mail.tsinghua.edu.cn}
\affiliation{State Key Laboratory of Low Dimensional Quantum Physics and Department of Physics, Tsinghua University, Beijing 100084, China}
\affiliation{Collaborative Innovation Center of Quantum Matter, Beijing, P.R. China}

\date{\today}

\begin{abstract}

{\bf In van der Waals heterostructures, the periodic potential from the Moir$\acute{e}$ superlattice can be used as a control knob to modulate the electronic structure of the constituent materials. Here we present a nanoscale angle-resolved photoemission spectroscopy (Nano-ARPES) study of transferred graphene/h-BN heterostructures with two different stacking angles of 2.4$^\circ$ and 4.3$^\circ$ respectively. Our measurements reveal six replicas of graphene Dirac cones at the superlattice Brillouin zone (SBZ) centers. The size of the SBZ and its relative rotation angle to the graphene BZ are in good agreement with Moir$\acute{e}$ superlattice period extracted from atomic force microscopy (AFM) measurements. Comparison to epitaxial graphene/h-BN with 0$^\circ$ stacking angles suggests that the interaction between graphene and h-BN decreases with increasing stacking angle.
}

~\\

{\bf keywords}: graphene/h-BN, van der Waals heterostructure, Moir$\acute{e}$ potential, nanoscale angle-resolved photoemission spectroscopy (nano-ARPES)%~\\

\end{abstract}

\maketitle

\section{Introduction}

\begin{figure*}[!htb]
\includegraphics[width=16.5 cm] {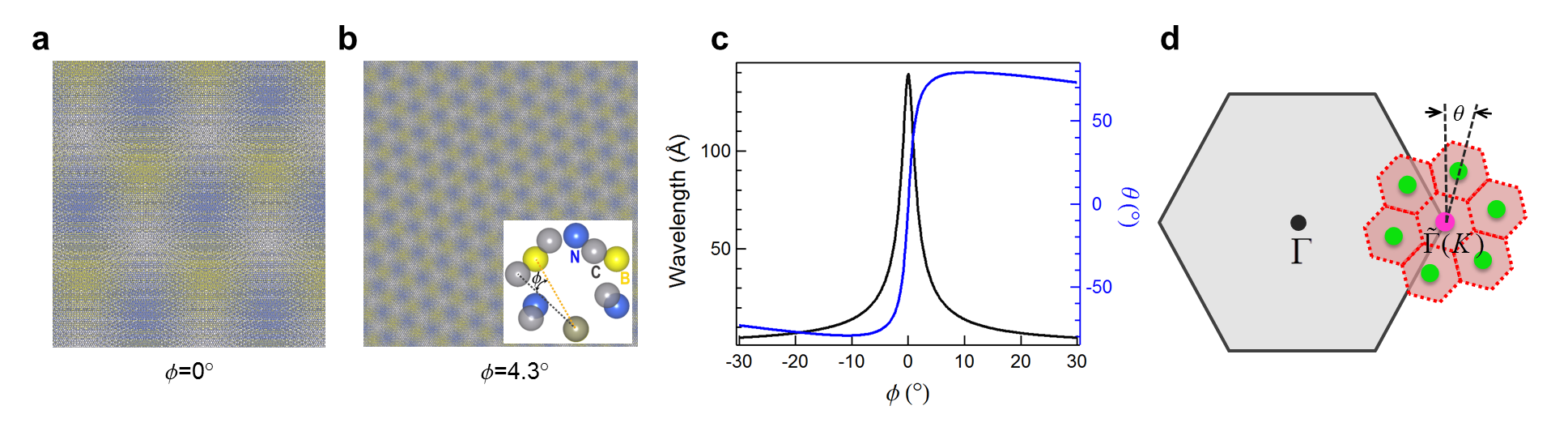}
\caption{{\bf Moir$\acute{e}$ pattern in graphene/h-BN heterostructure.} (a) Schematic drawing of the Moir$\acute{e}$ pattern in graphene/h-BN heterostructure with 0$^{\circ}$ stacking angle. (b) Schematic drawing of the Moir$\acute{e}$ pattern in graphene/h-BN heterostructure with 4.3$^{\circ}$ stacking angle. The inset schematically shows the relative stacking angle between graphene and h-BN. (c) Calculated Moir$\acute{e}$ pattern period $\lambda$ and rotation angle $\theta$ as a function of the stacking angle $\phi$. (d) Schematic drawing of the Brillouin zones of graphene and Moir$\acute{e}$ pattern. The $\Gamma$ (black circles), $\tilde{\Gamma}$ (K) (pink circles) points of graphene Brillouin zone and the six nearest SBZ centers (green circles) are shown.}
\label{Figure:Moire}
\end{figure*}

Van der Waals heterostructures are designed heterostructures made by stacking different two dimensional materials which are held together through weak van der Waals interaction \cite{GeimVDW}.  Such heterostructures not only broaden the range of materials that can be assembled and investigated, but also provide an important playground for discovering new properties different from the constituent materials and for realizing new quantum phenomena. In the past few years, graphene/h-BN has emerged as a model van der Waals heterostructure. It is an ideal system for making high quality graphene devices, with reduced ripples and higher mobility \cite{HoneNano2010, LeRoyNatureMater11}. Moreover, the moir$\acute{e}$ superlattice induced by the lattice mismatch and crystal orientation can significantly modify the electronic properties of graphene, leading to various novel quantum phenomena including the self-similar Hofstadter butterfly states \cite{GeimNature13, KimNature13, GBNgap,GeimNaturePhys2014} and topological currents \cite{GeimSci14}. There are also major changes in the electronic properties, e.g. emergence of second-generation Dirac cones (SDCs) \cite{GeimNature13, KimNature13, GBNgap,GeimNaturePhys2014}, renormalization of the Fermi velocity \cite{LouieNaturePhys08, FalkoMiniband, JvdBPRB12, Guinea},  gap opening at the Dirac point\cite{DFTgap, GBNgap,DGGPRL13, GeimSci14,NovoselovNphys2014} and gate-dependent pseudospin mixing \cite{FWangGate}. Although graphene/h-BN heterostructure has been studied by various transport and STM measurements \cite{GeimNature13, KimNature13, GBNgap,GeimNaturePhys2014,LeRoyNaturePhys12}, direct angle-resolved photoemission spectroscopy (ARPES) measurements of the modulated electronic band structure have been missing.

\begin{figure*}
\includegraphics[width=16.5 cm] {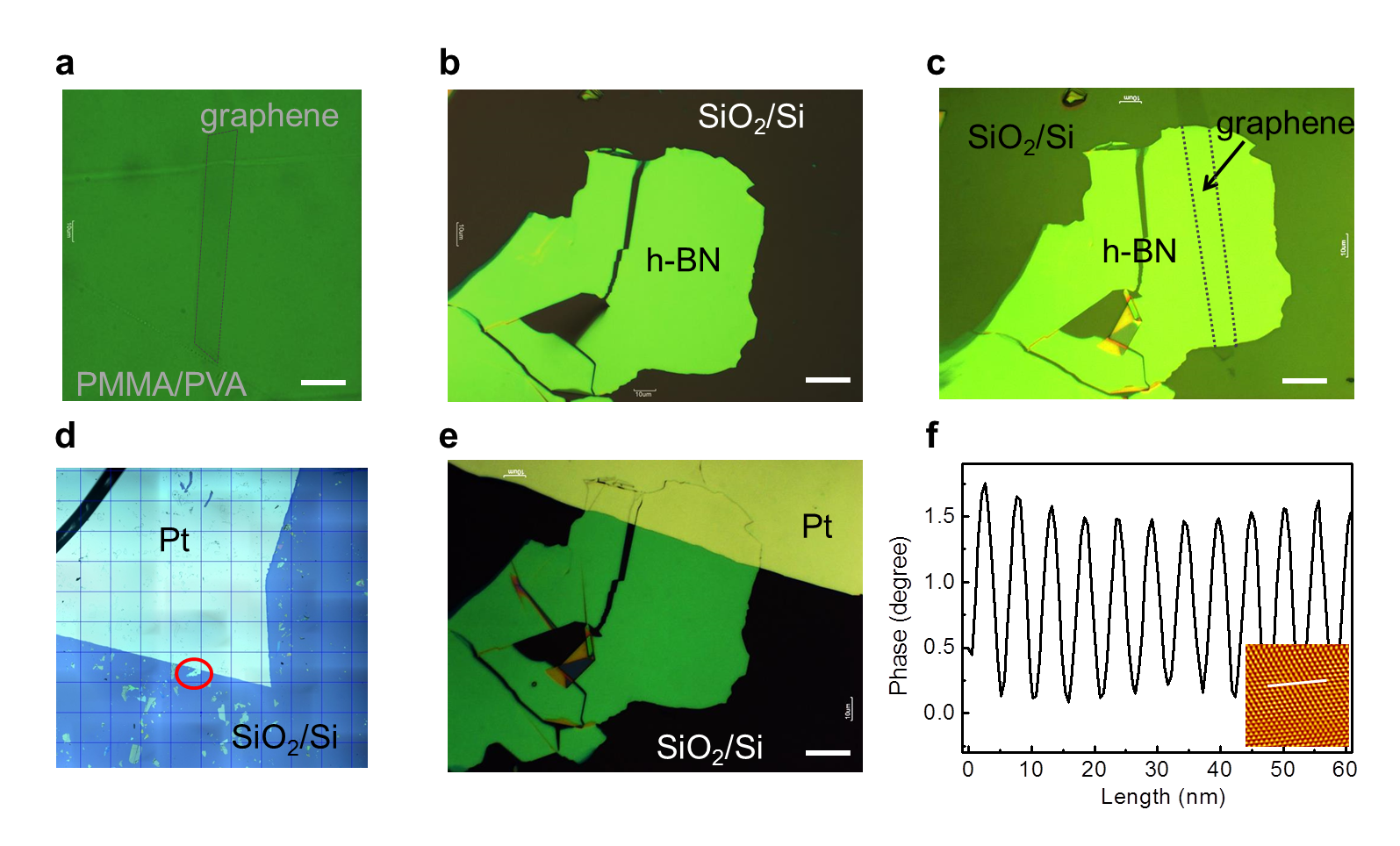}
\caption{{\bf Preparation of transferred graphene/h-BN heterostructure with 2.4$^{\circ}$ stacking angle.} (a) Exfoliated graphene flake on a suspended PMMA/PVA film. (b) Exfoliated h-BN flake on SiO$_2$/Si substrate. (c) Transferred graphene/h-BN heterostructure. (d) Optical image of the whole sample after deposition of Pt electrode. The location of target graphene/h-BN flake is highlighted by red circle. The square size is $500 {\mu}m {\times} 500 {\mu}m$. (e) Zoom-in of graphene/h-BN sample being measured. The white scale bars shown in above panels are 20 ${\mu}m$. (f) Phase profile along the white line in the inset AFM image. The inset shows the AFM image of the moir$\acute{e}$ pattern after high-pass-filtered inverse fast Fourier transformation.}
\label{Figure:sample24}
\end{figure*}

\begin{figure*}
\includegraphics[width=16.5 cm] {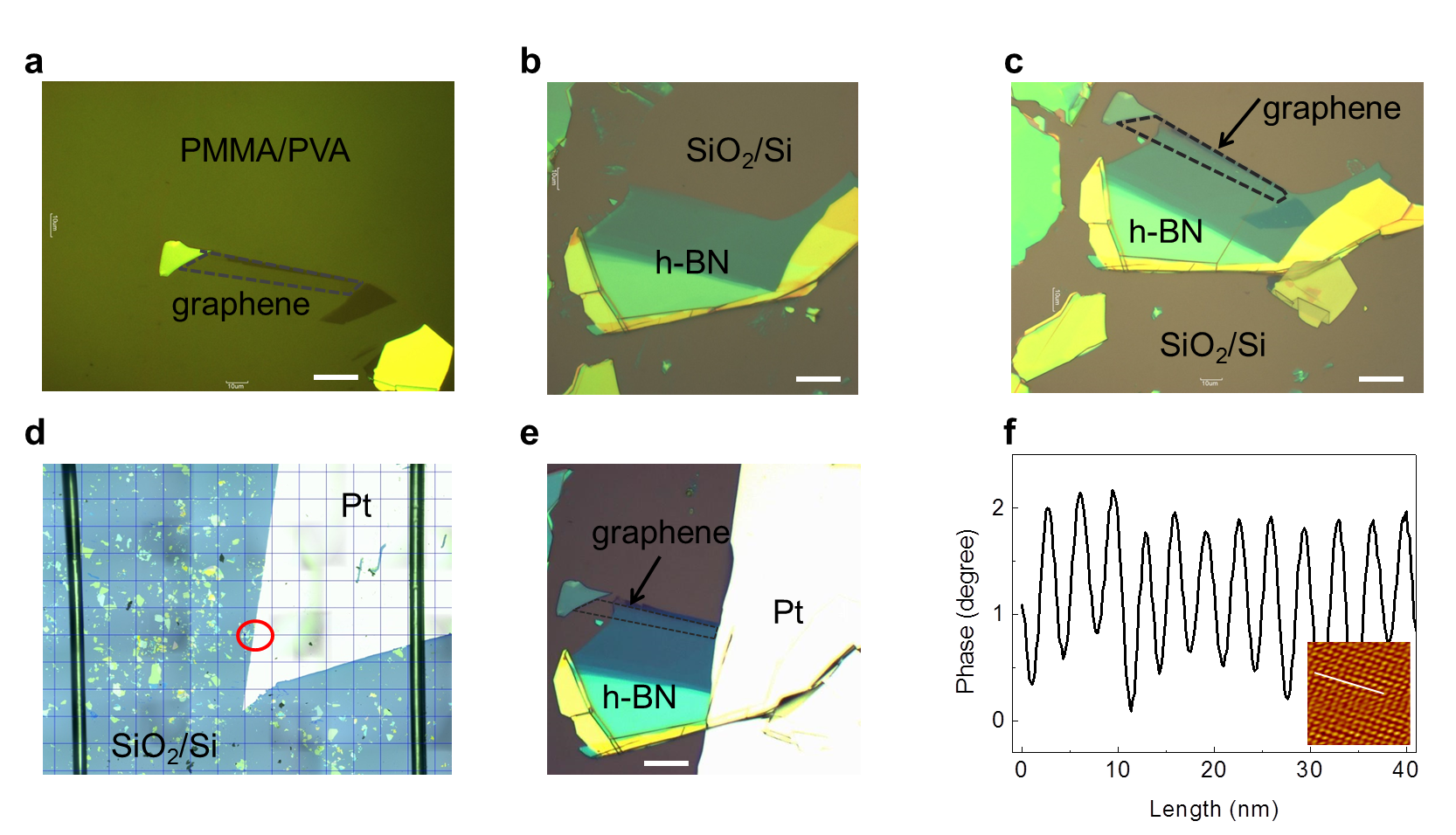}
\caption{{\bf Preparation of transferred graphene/h-BN heterostructure with 4.3$^{\circ}$ stacking angle.} (a) Exfoliated graphene flake on a suspended PMMA/PVA film. (b) Exfoliated h-BN flake on SiO$_2$/Si substrate. (c) Transferred graphene/h-BN heterostructure. (d) Optical image of the whole sample after deposition of Pt electrode. The location of target graphene/h-BN flake is highlighted by red circle. The square size is 500 ${\mu}$m ${\times}$ 500 ${\mu}$m. (e) Zoom in of graphene/h-BN sample being measured. The white scalebars shown in above panels are 20 ${\mu}$m. (f) Phase profile along the white line in the inset AFM image. The inset shows the AFM image of the moir$\acute{e}$ pattern after high-pass-filtered inverse fast Fourier transformation.}
\label{Figure:sample21}
\end{figure*}

Hexagonal Boron nitride (h-BN) shares similar honeycomb lattice structure with graphene, yet with a $\delta$ $\approx$1.8$\%$ larger lattice constant.  The breaking of the inversion symmetry by distinct boron and nitrogen sublattices  leads to a large band gap (5.97 eV) in the $\pi$ band, which is in sharp contrast to the gapless Dirac cones in graphene. By placing graphene atop h-BN, the different lattice constant and relative stacking angle $\phi$ between graphene and h-BN lead to moir$\acute{e}$ pattern. The moir$\acute{e}$ pattern periodicity is $$\lambda = \frac{(1+{\delta})a}{\sqrt{2(1+{\delta})(1-cos{\phi})+{\delta}^2 }} \quad (1) $$ where a is the lattice constant of graphene. The relative rotation angle $\theta$ of the moir$\acute{e}$ pattern with respect to the graphene lattice is given by
$$tan{\theta}= \frac{sin{\phi}}{(1+{\delta})-cos{\phi}} \quad  (2) \quad $$
Figure 1(a,b) shows the dependence of the moir$\acute{e}$ periodicity and rotation angle on $\phi$. The moir$\acute{e}$ periodicity quickly decreases with increasing $\phi$ and the rotation angle of the  moir$\acute{e}$ pattern also strongly depends on $\phi$. For example, when $\phi$ changes from 0$^{\circ}$ to 4.3$^{\circ}$, the moir$\acute{e}$ periodicity $\theta$ changes from 14 nm to 3.2 nm, and the relative angle changes from 0$^\circ$ to 74.5$^{\circ}$. The superlattice Brillouin zones (SBZs) are also rotated with respect to the Brillouin zone of graphene due to the rotated moir$\acute{e}$ pattern which is shown in Fig.\ref{Figure:Moire}(d) schematically. One expected result of the induced moir$\acute{e}$ superlattice potential is the formation of first-generation Dirac cones (FDCs) which occur at the same energy level as the original Dirac cone (ODC) yet translated by the reciprocal lattice vector of the moir$\acute{e}$ pattern G = $\frac{2}{\sqrt{3}}$ 2$\pi$/$\lambda$.   Furthermore, due to the induced moir$\acute{e}$ potential, second-generation Dirac cones (SDCs) can emerge at energies different from the ODC \cite{GeimNaturePhys2014,JvdBPRB12,MoonPRB14,FalkoMiniband}, and they are critical for the realization of self-similar Hofstadter bufferfly states under applied magentic field.

\begin{figure*}[!htb]
\includegraphics[width=16.5 cm] {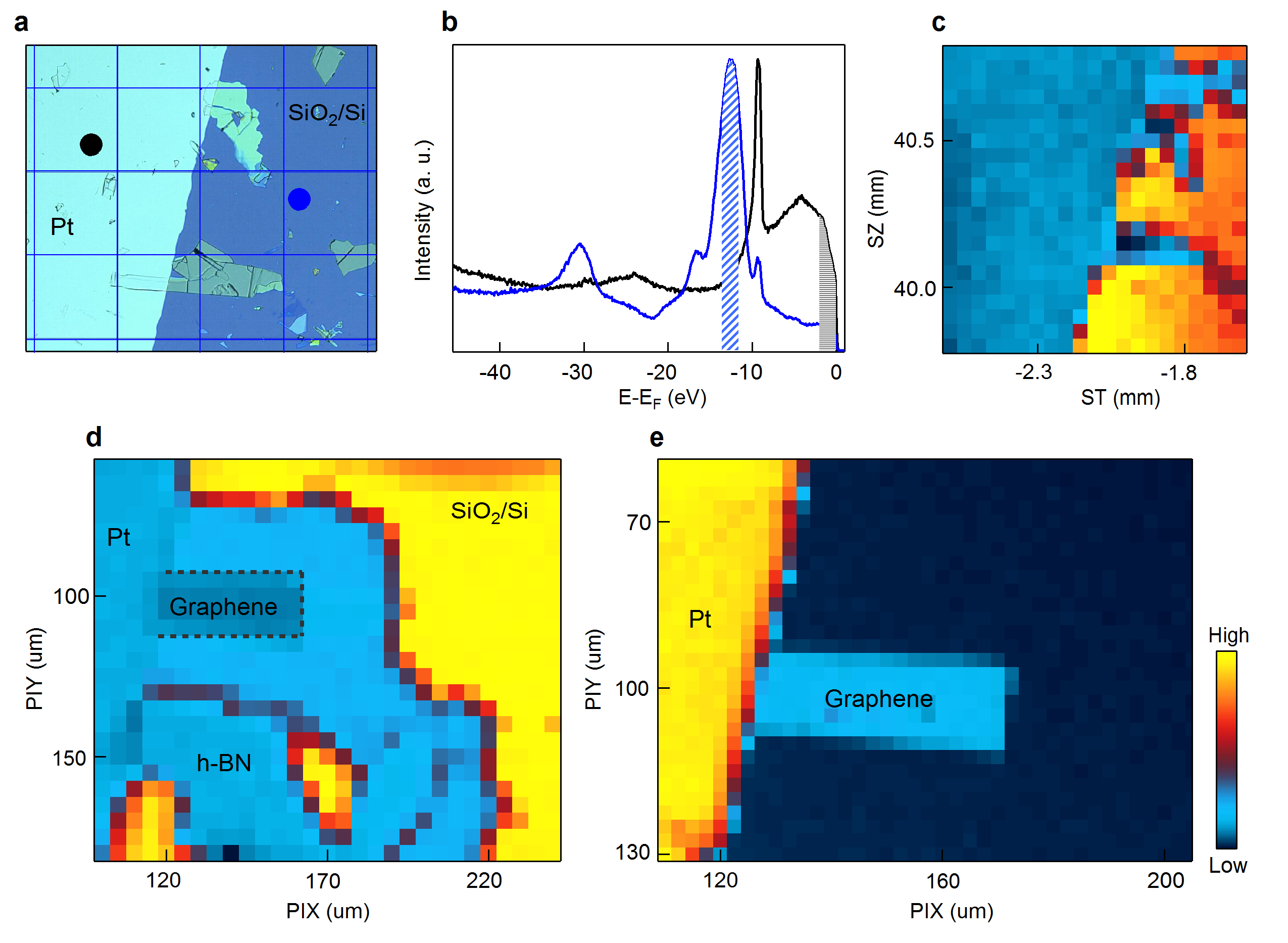}
\caption{{\bf Identification of graphene/h-BN heterostructure with 2.4$^{\circ}$ stacking angle from angle-integrated intensity maps.} (a) Optical image of sample \#1. The square size is 200 $\mu$m ${\times}$ 200 $\mu$m. (b) Two characteristic angle-integrated intensity curves from the Pt(black) and SiO$_2$/Si(blue) as indicated in (a). (c) Spatial map obtained by integrating the intensity of the blue shadow area shown in (b). (d) Zoom-in of sample area from (c) with finer steps. (e) Intensity map around the graphene/h-BN flake by integrating the grey shadow area near the Fermi energy shown in (b). }
\label{Figure:XPS}
\end{figure*}

Graphene/h-BN heterostructures can be prepared by directly growing epitaxial graphene on h-BN substrate using plama-enhanced chemical vapor deposition (PE-CVD) \cite{ZhangGY} or by transferring exfolicated graphene onto the h-BN substrates \cite{HoneNano2010,LeRoyNatureMater11}. While the stacking angle in PE-CVD grown samples is fixed to 0$^{\circ}$ \cite{ZhangGY}, the stacking angle in transferred graphene/h-BN samples can be carefully aligned and is widely tunable \cite{LeRoyNaturePhys12, NovoselovNphys2014, Coryscience}. With increasing stacking angle, the interaction between graphene and h-BN is expected to become weaker and commensurate-incommensurate transition has been reported \cite{NovoselovNphys2014}.  The PE-CVD heterostructures with 0$^{\circ}$ stacking angle and sample size up to a few hundred micrometers ($\mu$m) have been recently studied by regular ARPES, and SDCs have been reported \cite{GBNCVD}. However, until now direct experimental results on the modulated band structure in graphene/h-BN with nonzero stacking angles have been missing. The challenge is related to the small sample size of a few micrometers ($\mu$m), which is much smaller than the typical ARPES beam size of 50-100 $\mu$m. By using nano-ARPES with beam size of $\sim$100 nm, we are able to obtain the electronic structure of transferred graphene/h-BN heterostructures with nonzero stacking angles for the first time. Here we report the electronic structure of such heterostructures with stacking angles of  $\sim$ 2.4$^{\circ}$ and $\sim$ 4.3$^{\circ}$ respectively, and reveal the FDCs induced by the moir$\acute{e}$ superlattice potential.

\section{Methods}

\begin{figure*}[!htb]
\includegraphics[width=16.5 cm] {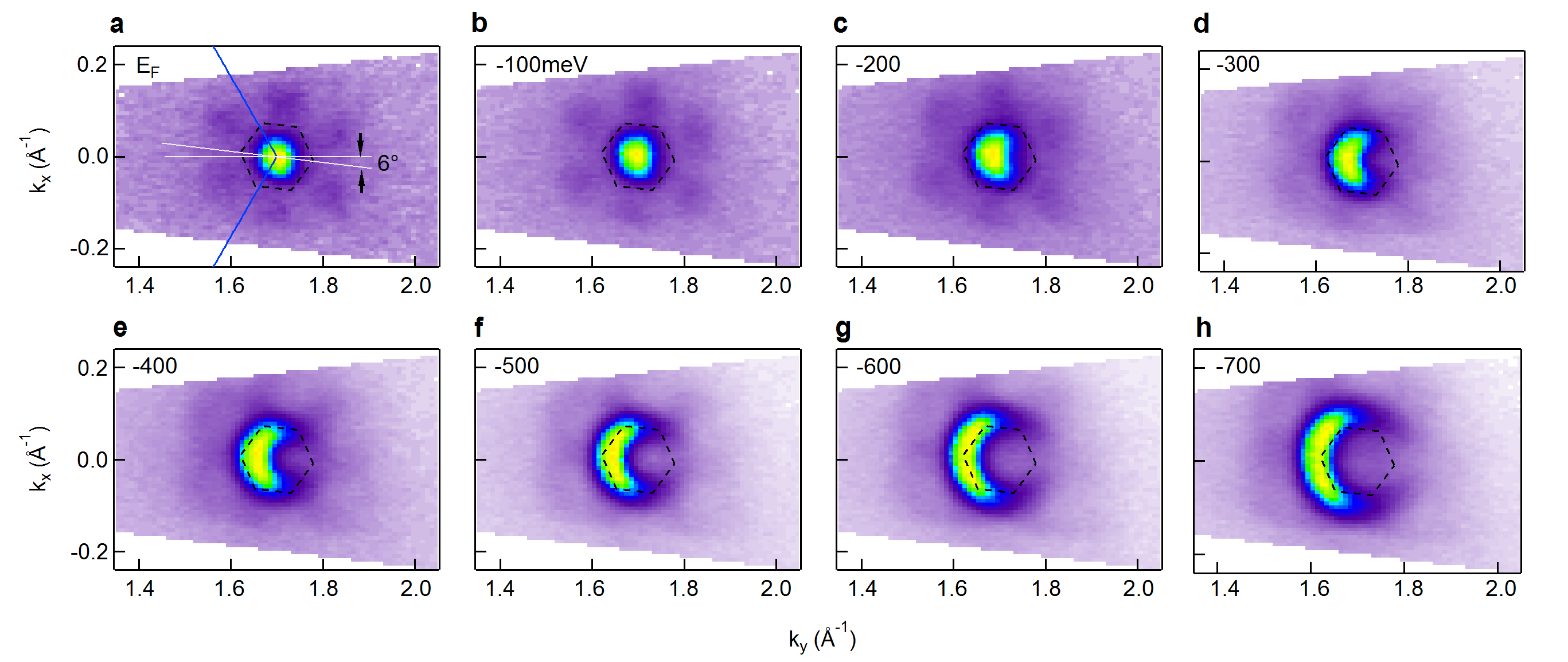}
\caption{{\bf First-generation Dirac cones in the intensity maps of sample \#1 with 2.4$^{\circ}$ stacking angle.} (a-h) Intensity maps at E$_F$, -100, -200, -300, -400, -500, -600 and -700 meV respectively. The Brillouin zones of graphene and moir$\acute{e}$ pattern are indicated by blue solid and black dashed lines.}
\label{Figure:FS24}
\end{figure*}

Transferred graphene/h-BN heterostructures were prepared using similar methods as reported \cite{HoneNano2010,LeRoyNatureMater11}. The entire process is shown in Figure \ref{Figure:sample24}. First, single layer graphene was exfoliated on polymethyl methacrylate(PMMA, MicroChem, A6, 950K)/polyvinyl alcohol(PVA, Sigma-Aldrich) stack (Fig.~\ref{Figure:sample24}(a)). Single crystals of h-BN were grown using the method described before \cite{hBNgrow} and h-BN flakes were exfoliated on a SiO$_2$/Si wafer (Fig.~2\ref{Figure:sample24}(b)).
Then the exfoliated graphene was transferred atop exfoliated h-BN flakes (Fig.\ref{Figure:sample24}(c)). Pt electrode was deposited on one side of SiO$_2$/Si substrate  (Fig.\ref{Figure:sample24}(d,e)) to avoid charging effect during ARPES measurements. Atomic Force Microscopy (AFM) measurements were performed to confirm the existence of moir$\acute{e}$ pattern.  The extracted moir$\acute{e}$ periodicity can also be used to determine the stacking angle between graphene and h-BN. Figure \ref{Figure:sample24}(f) shows the extracted phase profile for sample \#1. The moir$\acute{e}$ periodicity is estimated to be $5.3\pm0.2$ nm and the stacking angle is 2.4$^{\circ}\pm 0.2^{\circ}$ calculated from equation (1). The optical images for sample \#2 during the sample preparation are shown in Fig.\ref{Figure:sample21}(a-e). AFM measurements were also performed to verify the existence of moir$\acute{e}$ pattern afterwards. Figure \ref{Figure:sample21}(f) shows the extracted phase profile from AFM image. The moir$\acute{e}$ pattern periodicity and the stacking angle are extracted to be $3.2\pm0.3$ nm and  4.3$^{\circ}\pm 0.3^{\circ} $  respectively.

\begin{figure*}[!htb]
\includegraphics[width=16.5 cm] {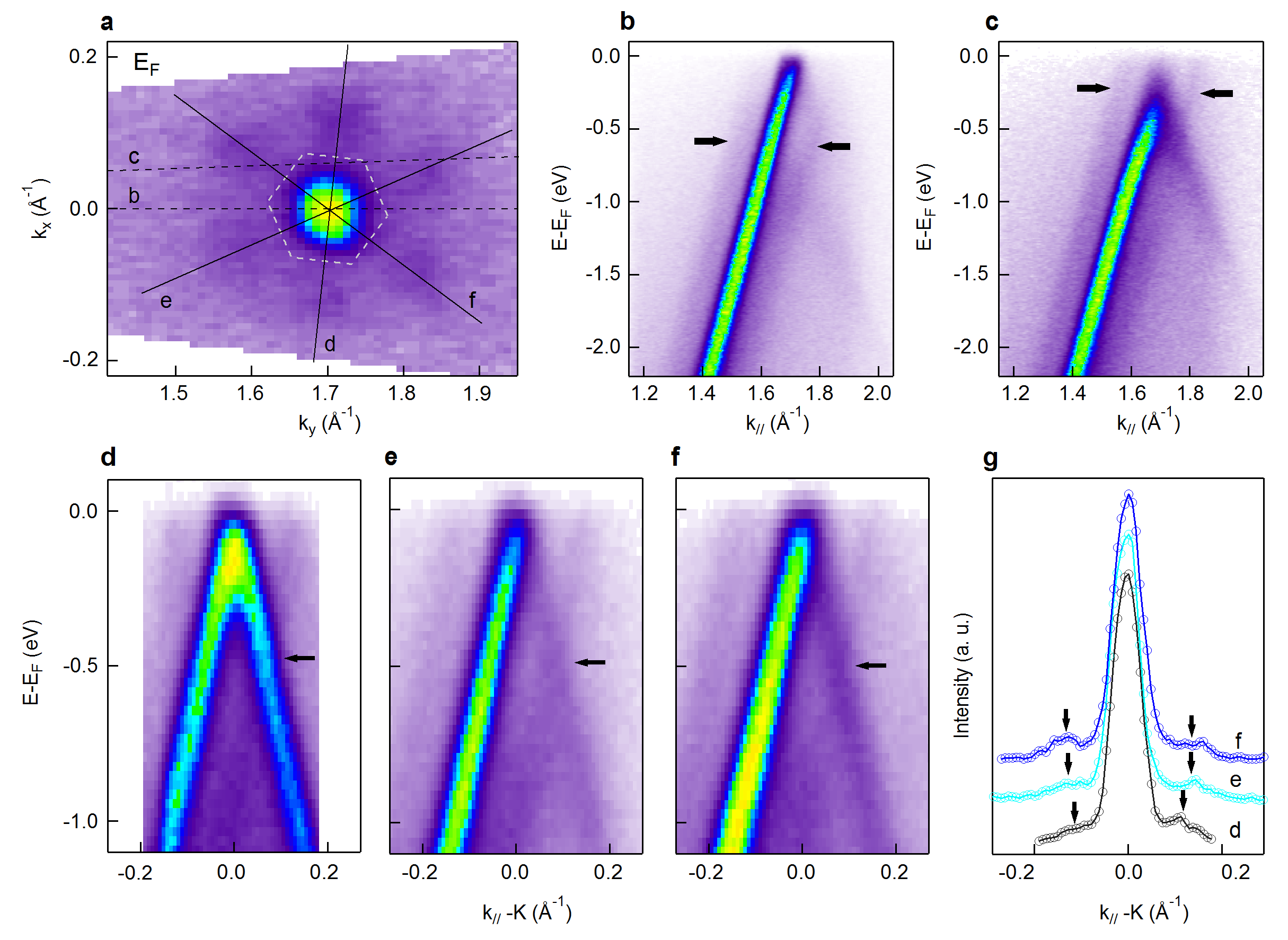}
\caption{{\bf Detailed band dispersions of sample \#1 with 2.4$^{\circ}$ stacking angle.} (a) Fermi surface map. The Brillouin zone of the moir$\acute{e}$ superlattice is indicated by grey dashed hexagon. The lines mark the position of cuts shown in b to f. (b-c) Band dispersions along the cuts shown in (a). The dispersions from FDCs are indicated by black arrows. (d-f) Band dispersions along the cuts across the ODC and FDCs as shown in (a). The intersection positions of ODC and FDCs are indicated by black arrows. (g) MDCs at E$_F$ from (d-f). The momentum positions of first-generation Dirac points are indicated by black arrows.}
\label{Figure:Cut24}
\end{figure*}

Nano-ARPES measurements were performed at the ANTARES beamline of the SOLEIL synchrotron. ANTARES beamline is equipped with two Fresnel zone plates (FZPs) to focus the beam size down to $\approx$ 120 nm.  There are two sets of motor systems to change the sample position, mechanical motors SZ and ST which have large motion range with $\approx$ 5 $\mu$m resolution and piezoelectric motors PIX and PIY which have smaller motion range and better spatial resolution $\approx$ 5 nm. Samples were mounted on a nanopositioning stage which was placed at the coincident focal point of the electron analyzer and the FZPs. ANTARES can operate in two modes, the imaging mode where the photoelectron spectra (angle integrated or angle resolved) are collected by changing the sample position to create two-dimensional images of electronic states of interest, and the spectroscopy mode where the detailed band dispersions are measured at fixed sample position. The data were recorded with Scienta R4000 analyzer with photon energy of 100 eV using horizontal linear polarized light. The samples were annealed at $\approx$ 250 $^{\circ}$C until clean dispersions were obtained. The samples are kept at 80 K with vacuum better than $2 \times 10^{-10}$ Torr during the ARPES measurements.

\section{Results and discussion}

 We first used the imaging mode to locate the small target graphene/h-BN flake on SiO$_2$/Si substrate for sample \#1 (red circle in Fig.\ref{Figure:sample24}(d)). The angle-integrated intensity curves show two typical spectra as shown in Fig.\ref{Figure:XPS}(b). The black curve with stronger intensity near E$_F$ is attributed to the Pt electron, and the blue curve with suppressed intensity near E$_F$  and a peak at $\approx$ -13 eV is attributed to the bare SiO$_2$/Si substrate \cite{SiO2}. By integrating the spectral weight of the blue shadow area which is characteristic of SiO$_2$/Si (Fig.\ref{Figure:XPS}(b)), the spatial map (Fig.\ref{Figure:XPS}(c)) shows similar shape with SiO$_2$/Si in the optical image (Fig.\ref{Figure:XPS}(a)). This confirms our assignment of the two types of spectra. The zoom-in intensity map around the graphene/h-BN region (Fig.\ref{Figure:XPS}(d)) allows to further distinguish different parts (Pt, graphene, h-BN and SiO$_2$/Si) clearly. Moreover, by integrating the spectral weight near the Fermi energy (grey shadow area in Fig.\ref{Figure:XPS}(b)) to have a better contrast between graphene/h-BN and bare h-BN, the graphene flake can be clearly identified (Fig.\ref{Figure:XPS}(e)).

\begin{figure*}[!htb]
\includegraphics[width=16.5 cm] {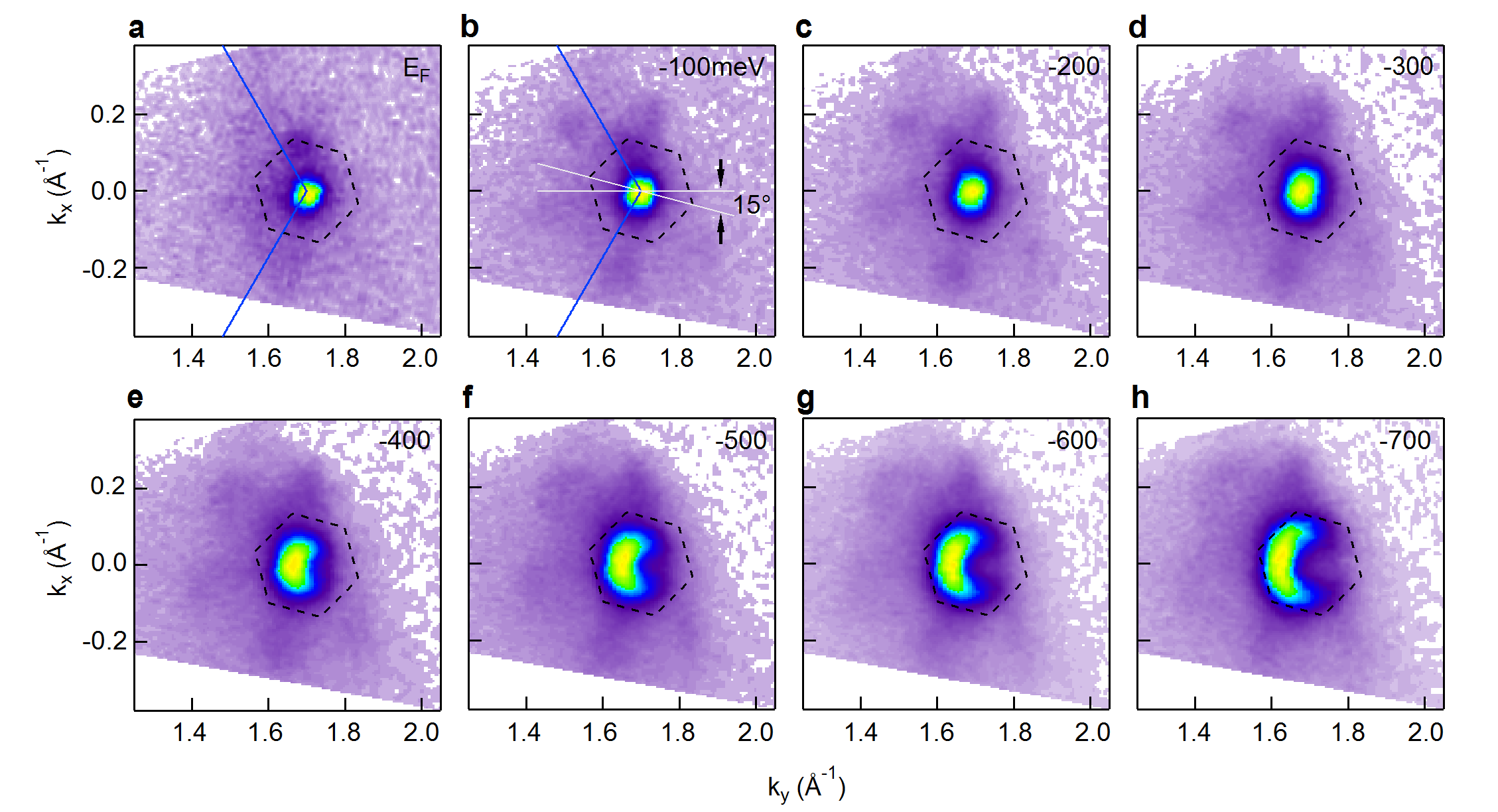}
\caption{{\bf First-generation Dirac cones shown in the intensity maps of sample \#2 with 4.3$^{\circ}$ stacking angle.} (a-h) Intensity maps at E$_F$, -100meV, -200meV, -300meV, -400meV, -500meV, -600meV and -700meV respectively. The Brillouin zones of graphene and moir$\acute{e}$ pattern are indicated by blue solid and black dashed lines. }
\label{Figure:FS21}
\end{figure*}

After locating target graphene/h-BN flake, we used the spectroscopy mode to probe the modulated band structure of transferred graphene/h-BN. Figure \ref{Figure:FS24} shows the intensity maps at energies from E$_F$ to -700 meV.  At E$_F$, six cloned FDCs are observed around the ODC, consistent with our previous ARPES studies on PE-CVD graphene/h-BN heterostructures \cite{GBNCVD}. The corresponding SBZ (black dashed hexagon) is rotated with respect to the Brillouin zone of graphene (blue solid line) by $\approx$ 6$^\circ$, which is consistent with the calculated  rotation angle for the moir$\acute{e}$ pattern  $\theta$ $\approx$ 66$^\circ$ (effectively 6$^\circ$ in the map due to the six fold symmetry) from the stacking angle $\phi$ $\approx$ 2.4$^\circ$ by equation (2). The size of the pockets  becomes larger with decreasing energy, consistent with the conical dispersions expected. Figure \ref{Figure:Cut24} further shows detailed cuts around the ODC and FDCs to reveal the modulated band structures. The dispersions from the FDCs are obvious on both sides of ODC in Fig.~\ref{Figure:Cut24}(b,c). Panels (d-f) show the dispersions along the cuts crossing the ODC and FDCs simultaneously. From the momentum distribution curves(MDCs) at the Fermi energy (Fig.~\ref{Figure:Cut24}(g)), we extract the momentum separation between the ODC and FDCs to be $0.134 \pm 0.02$ ${\AA}^{-1}$ which is consistent with the moir$\acute{e}$ pattern periodicity inferred from AFM measurements. Using a Fermi velocity of 7.37 eV$\cdot{\AA}$ (1.13$\times10^{6}$ m/s), the intersection point for 2.4 $^\circ$ heterostructure is estimated to be at -493 meV. From ARPES measurements, the crossing point is measured to be at -490 meV (pointed by black arrows in Fig.~\ref{Figure:Cut24}(d-f)), in agreement with the estimation, however in contrast with our previous studies on 0$^{\circ}$ aligned graphene/h-BN \cite{GBNCVD}, no obvious signatures of SDCs are observed from the constant energy maps and band dispersions. This is possibly due to the weaker interaction between graphene and h-BN when increasing the stacking angle between graphene and h-BN.

\begin{figure*}
\includegraphics[width=16.5 cm] {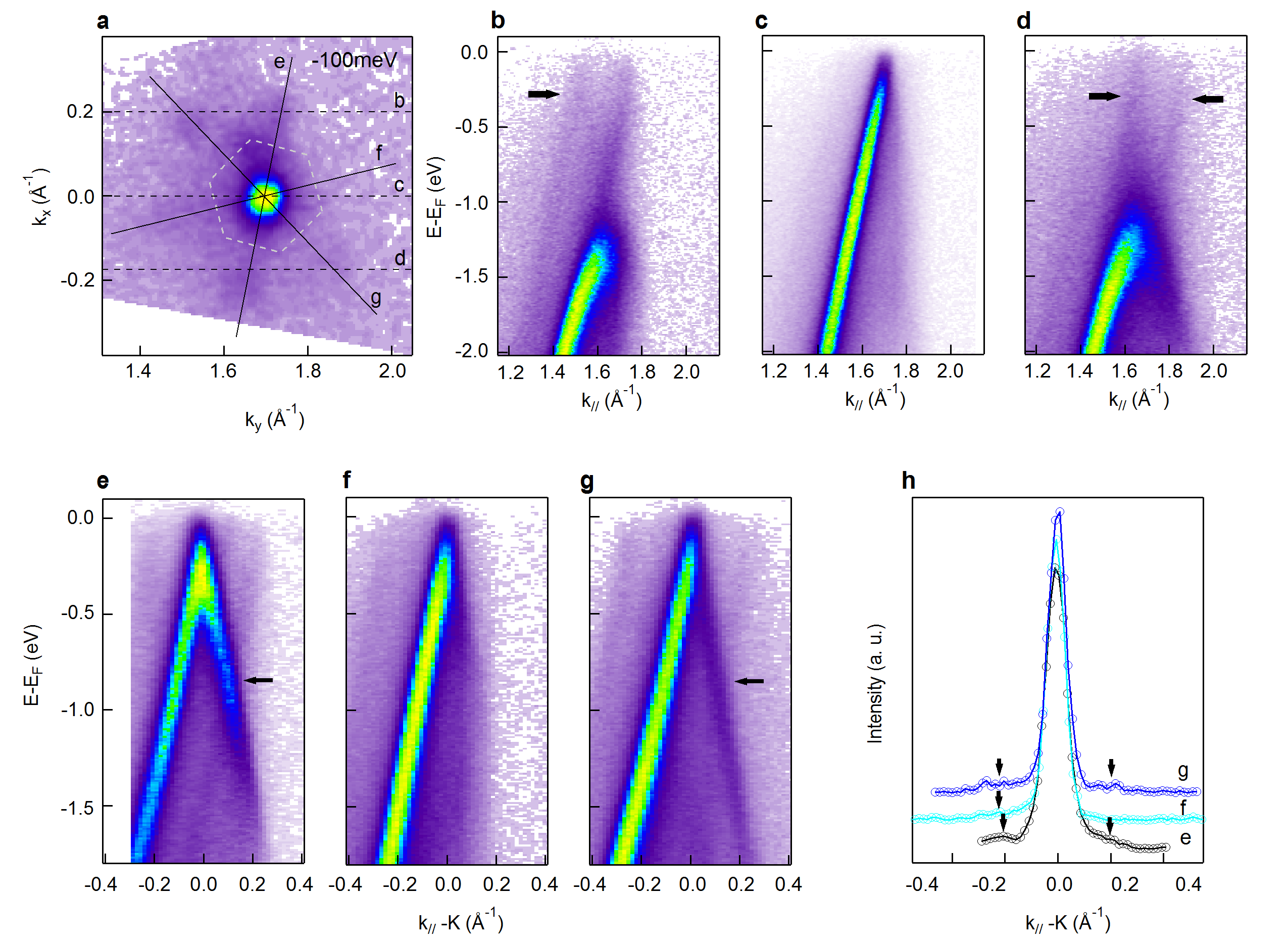}
\caption{{\bf Detailed band dispersions of sample \#2 with 4.3$^{\circ}$ stacking angle.} (a) Constant energy map at E$_F$-100 meV. The Brillouin zone of moir$\acute{e}$ superlattice is indicated by grey dashed hexagon. (b-d) Band dispersions along the cuts shown in (a). The dispersions from FDCs are indicated by black arrows. (e-g) Band dispersions along the cuts across the ODC and FDCs as shown in (a). (h) MDCs at E$_F$-100 meV for (e-g). The momentum positions of first-generation Dirac points are indicated by black arrows.}
\label{Figure:Cut21}
\end{figure*}

 We applied the same method to locate the target graphene/h-BN flake for sample \#2. Then spectroscopy mode was used to probe the detailed band structure. Figure \ref{Figure:FS21} shows the intensity maps at from E$_F$ to - 700 meV. The signal from FDCs is weak at E$_F$ and becomes more clear at  -100 meV. This could be attributed to even weaker interaction between graphene and h-BN with an increasing stacking angle. The corresponding SBZ (black dashed hexagon) is rotated by $\approx$15$^\circ$ with respect to the Brillouin zone of graphene (blue solid lines in Fig.\ref{Figure:FS21}(b)), which is consistent with the calculated moir$\acute{e}$ pattern rotation angle of 74.5$^\circ$ from the stacking angle of 4.3$^\circ$. Figure.\ref{Figure:Cut21} shows the detailed modulated band dispersions  around the ODC and FDCs. The dispersions from the FDCs are weak but still detectable on both sides of the ODC (Fig.\ref{Figure:Cut21}(b-d)). Panel (e-g) shows the dispersions along the cuts crossing the ODC and FDCs simultaneously. The momentum separation between ODC and FDCs is estimated to be 0.23$\pm$0.04 ${\AA}^{-1}$ from the MDCs at -100 meV (Fig.\ref{Figure:Cut21}(h)), consistent with the moir$\acute{e}$ pattern period extracted from AFM measurements. The ODC and FDCs intersect at around -840 meV, slightly deeper than the -770 meV calculated by moir$\acute{e}$ pattern period. The difference is possibly attributed to the error bar from AFM measurement. Similar to the graphene/h-BN sample \#1 with stacking angle of the 2.4$^{\circ}$, no obvious signals of SDCs are observed from the constant energy maps or band dispersions.

\section{Conclusions}

We report direct experimental results on the modulated band structure in transferred graphene/h-BN heterostructures with stacking angles of 2.4$^\circ$ and 4.3$^\circ$ respectively. We observed replicas of Dirac cones translated by the reciprocal lattice vectors of the Moir$\acute{e}$ superlattice from the graphene K point. With variable stacking angle between graphene and h-BN, the size and relative angle of SBZ can be tuned. Unlike previous studies on PE-CVD grown graphene/h-BN samples with 0$^\circ$ stacking angle, no obvious signatures of SDCs are observed from the modulated band structure. This is possibly due to weaker interaction between graphene and h-BN at large stacking angle. It has been reported that at small stacking angle, there are regions of commensurate graphene stretched to fit the lattice of h-BN substrate, and there is a crossover from commensurate to incommensurate states between $\phi$ = 0 and $\phi$ $\approx$ 1.5 $^\circ$ \cite{NovoselovNphys2014}.  Such commensurate to incommensurate transition may also explain the large variations of gaps measured on graphene samples with different stacking angles. Another possibility is that PE-CVD graphene/h-BN samples may have stronger interaction between graphene and h-BN with 0$^\circ$ stacking angle compared to transferred graphene/h-BN samples with tunable stacking angles.  More experiments to reveal the atomic structure at the interface with different stacking angles and their correlation with the electronic structures are important to further understand the difference.

{\bf Acknowledgements}
This work is supported by the National Natural Science Foundation of China (Grant No.~11274191, 11334006, and 11427903), Ministry of Science and Technology of China (Grant No.~2015CB921001) and Tsinghua University Initiative Scientific Research Program (2012Z02285). The Synchrotron SOLEIL is supported by the Centre National de la Recherche Scientifique (CNRS) and the Commissariat $\grave{a}$  l'Energie Atomique et aux Energies Alternatives (CEA), France.

{\bf Author Contributions}
S.Z. designed the research project. E.W., G.W., C.C., J.A., M.C.A. and S.Z. performed the ARPES measurements and analyzed the ARPES data.  G.C., X.L., G.Z and Y.Z. prepared the graphene samples. E.W. and S.Z. wrote the manuscript, and all authors commented on the manuscript.

{\bf Competing financial interests}
The authors declare no competing financial interests.

\end{document}